\documentclass[aps,prl,twocolumn,showpacs]{revtex4}
\usepackage{amssymb}
\usepackage{amsmath, amsthm}

\begin{document}


\title{The Hawking temperature of expanding cosmological black 
holes}


\author{Valerio Faraoni}
\email[]{vfaraoni@ubishops.ca}
\affiliation{Physics Department, Bishop's University\\
Sherbrooke, Qu\'ebec, Canada J1M~0C8
}


\begin{abstract} 
In the context of a debate on the correct expression 
of the Hawking temperature of a cosmological black hole, we 
show that the correct expression in terms of the  
Hawking-Hayward quasi-local energy  $m_H$ of the hole is 
$T=\left( 8\pi 
m_H(t) \right)^{-1}$. This expression holds for  comoving 
black holes and agrees  with a recent proposal 
by Saida, Harada, and Maeda.
\end{abstract}


\maketitle

\section{Introduction}
\setcounter{equation}{0}

Recently, the power of Hawking radiation emitted by a 
cosmological black hole was computed by Saida, Harada, and 
Maeda 
\cite{SHM}. In this work,  the black hole is not the usual 
asymptotically flat spacetime but lives instead in an 
asymptotically Friedmann-Lemaitre-Robertson-Walker (FLRW) 
universe, which is taken to have flat spatial sections for 
simplicity. In Ref.~\cite{SHM}, two exact solutions of the 
Einstein 
equations describing such systems are considered: the 
Einstein-Straus vacuole \cite{EinsteinStraus} with a central 
non-expanding black hole, and the recent Sultana-Dyer solution 
\cite{SD} describing a perfectly comoving black hole embedded 
in a dust-dominated FLRW universe  (see Refs. 
\cite{otherblackholes} for 
other dynamical black hole solutions and 
\cite{Kothawalaetal,Criscienzoetal} for the thermodynamics of 
dynamical black hole horizons).

The analysis of Saida, Harada, and Maeda \cite{SHM} delivers 
two main results: for 
the Einstein-Straus black hole 
\cite{EinsteinStraus}, 
which is not accreting, thermal  radiation of quantum 
particles is suppressed by a 
factor coming from the expansion of the boundary between the 
local (static) black hole exterior and the expanding FLRW 
universe. This phenomenon is interpreted in analogy with 
radiation 
from an accelerated mirror, and we will not be concerned with 
it here. The second result, upon which we focus, pertains to  
the 
second exact solution studied in \cite{SHM}, {\em i.e.}, the 
Sultana-Dyer cosmological black hole \cite{SD}. This solution 
is obtained by conformally transforming the Schwarzschild 
metric
\begin{eqnarray}
ds^2_{Schw} &= & -\left( 1-\frac{2m}{r} \right)dt^2 +
\left( 1-\frac{2m}{r} \right)^{-1}dr^2 \nonumber \\
&&\nonumber \\
&+& r^2 \left( 
d\theta^2+\sin^2\theta\, d\varphi^2 
\right)=g_{ab}^{(Schw)}dx^adx^b 
\end{eqnarray}
according to
\begin{equation}\label{}
g_{ab}^{(Schw)}\longrightarrow g_{ab}^{(SD)}=\Omega^2 
g_{ab}^{(Schw)}
\end{equation}
and choosing the conformal factor $\Omega=a(t)$, the scale 
factor of a spatially flat FLRW metric
\begin{equation}\label{}
ds^2_{FLRW}=-dt^2+a^2(t)\left[ dr^2+r^2
\left( d\theta^2+\sin^2\theta\, d\varphi^2 \right)\right]
\end{equation}
with the particular choice $a(t)=a_0t^{2/3}$ (the scale factor 
of a dust-dominated universe). The explicit goal of Sultana and 
Dyer is to turn the Schwarzschild global timelike Killing field 
$\xi^c$ into a conformal Killing field (which happens for 
$\xi^c\nabla_c \Omega \neq 0$) which generates a conformal 
Killing horizon (the dynamical black hole horizon). The 
Sultana-Dyer metric can be written in various coordinate 
systems; for example,
\begin{eqnarray}
&& ds^2_{SD}=a^2(\eta)\left[ -d\eta^2 +dr^2 + r^2 
\left( d\theta^2+\sin^2\theta\, d\varphi^2 \right) \right. 
\nonumber \\
&&\nonumber \\
&& \left. + \frac{2m}{r} \left( d\eta +dr \right)^2 \right] 
\;,\label{SD1}
\end{eqnarray}
where $m=$const.$>0$, $\eta$ is the conformal time, and $r$ 
is an areal (Schwarzschild-like) radial coordinate. To see 
that~(\ref{SD1}) is conformal to the Schwarzschild line element, 
one performs the coordinate transformation $ \eta \equiv 
t+2m\ln \left( \frac{r}{2m}-1 \right)$ which turns (\ref{SD1}) 
into
\begin{eqnarray}
&& ds^2_{SD} = a^2( \eta )\left[ -\left(1-\frac{2m}{r} \right) 
d \eta^2  
+ \left(1-\frac{2m}{r} \right)^{-1} dr^2 \right. \nonumber \\
&&\nonumber \\
&& \left. +r^2 
\left( d\theta^2+\sin^2\theta\, d\varphi^2 \right) 
\right] \;.
\end{eqnarray}
It is relevant \cite{FJ} that the Sultana-Dyer solution can be 
seen as a generalization of the McVittie metric 
\cite{McVittie} describing a 
point particle embedded in a cosmological background. For a 
spatially flat FLRW background this is given by \cite{McVittie}
\begin{eqnarray}
ds^2_{McVittie} &= & 
- \frac{ \left( 1-\frac{m(t)}{ 2\bar{r} } \right)^2 }{
\left( 1+\frac{m(t)}{2\bar{r}} \right)^2}  dt^2
+ a^2(t) \left( 1+ \frac{m(t)}{2\bar{r}} \right)^4 \nonumber \\
&& \nonumber \\
& \cdot & \left[ 
d\bar{r}^2 + \bar{r}^2 \left( 
d\theta^2+\sin^2\theta\, d\varphi^2 
\right) \right] \;,\label{McVittiemetric} 
\end{eqnarray}
where $\bar{r}$ is the isotropic radius defined by 
\begin{equation}
r=\bar{r}\left( 1+\frac{m}{2\bar{r} }\right)^2 \;,
\end{equation}
$a(t)$ is the scale factor of the 
background FLRW universe, and $m(t)$ is a function of 
the comoving time $t$ related to the physical mass of the 
central object and determined by the McVittie condition
\begin{equation}\label{McVittiecondition}
\frac{\dot{m}}{m}+\frac{\dot{a}}{a}=0 \;,
\end{equation}
where an overdot denotes differentiation with respect to the 
comoving time. 
This condition is equivalent to $G_0^1=0$ (where $G_{cd}$ is 
the Einstein tensor) and to vanishing component $T_0^1$ of the 
energy-momentum tensor of the cosmic fluid. The physical 
meaning is that no accretion onto the central object occurs 
and, as  a consequence, the central object's radius stays 
constant, 
as discussed in Ref.~\cite{FJ}. This dictates the form of 
$m(t)=$constant$/a(t)$. The McVittie metric 
(\ref{McVittiemetric}) can not be interpreted as describing a 
cosmological black hole because at the putative horizon 
$\bar{r}=m/2$ (or $ r=2m$) the pressure of the cosmic fluid 
and the Ricci curvature $R$ diverge 
\cite{Sussman,Ferrarisetal,NolanCQG1}. An important 
exception is the Schwarzschild-de Sitter solution which is a 
special case of the metric (\ref{McVittiemetric}) and describes 
a genuine static black hole embedded in de Sitter space 
\cite{NolanCQG1}. 

The Sultana-Dyer solution corresponds to a generalized McVittie 
metric (\ref{McVittiemetric}) in which the condition 
(\ref{McVittiecondition}) is dropped and the central black hole 
is 
allowed to (radially) accrete the surrounding cosmic fluid. The 
Sultana-Dyer choice of $m(t)=$constant$ \equiv m_0$ is, in this 
respect, the simplest. The Sultana-Dyer solution does not 
satisfy the energy conditions; indeed, the energy density of 
the cosmic fluid even becomes negative, and the flow becomes 
superluminal, near the horizon at late times \cite{SD}. Other 
solutions with similar problems have been presented in 
Refs.~\cite{McClureDyerCQG, McClureDyerGRG}, and new solutions 
have been found in Ref.~\cite{FJ}; some of the latter have 
positive energy density everywhere on the spacetime manifold 
but still suffer from the superluminal flow problem due to the 
oversimplified model of ``rigid'' accretion.  However, they are 
not restricted to the special form $a(t)=a_0t^{2/3}$ but hold 
for general scale factors.

The Hawking-Hayward \cite{Hawking, Hayward} quasi-local 
energy 
of the McVittie, Sultana-Dyer, and new solutions of 
the form (\ref{McVittiemetric}) is \cite{NolanCQG1, FJ}
\begin{equation}
m_H(t)=a(t) \, m(t) \;;
\end{equation}
this should be regarded as the physical mass of the central 
object or black hole, where applicable, as opposed to the 
function $m(t)$ (which has led to misleading or incorrect 
statements in the literature \cite{McClureDyerCQG, GaoZhang}) 
or 
to the Misner-Sharp mass \cite{SHM} which is not particularly 
illuminating. In terms of $m_H$, the McVittie condition 
(\ref{McVittiecondition}) says that the Hawking mass stays 
constant  ($\dot{m}_H=0$, no accretion), while the Sultana-Dyer 
solution has 
$m_H(t)=m_0 a(t)$, {\em i.e.}, is ``perfectly comoving''. This 
is not 
an abuse of terminology because the physical mass $m_H$ is 
related to the physical size of the horizon by $r=2m_H 
=a(t)m_0$, which is reminiscent of the expression of the 
Schwarzschild 
radius $r=2m$. This relation comes from the definition of the 
areal radius $r\equiv \sqrt{ \frac{ {\cal A}}{4\pi}}$, where 
${\cal A}$ is the proper area of the horizon $\Sigma$,
\begin{equation}  \label{area}
{\cal A}= \int\int d\theta\, d\varphi \,\, \sqrt{ 
g_{\Sigma}}=16\pi a^2m^2 =16\pi m_H^2
\end{equation}
and $g_{\Sigma}$ is the determinant of the restriction 
$g_{ab}^{( \Sigma)}$ of the metric to this surface. 
Eq.~(\ref{area}) tells us that the  surface 
$ \bar{r}=m/2$ in the McVittie geometry (including the case of 
the Schwarzschild-de Sitter black hole) does not expand, while 
the Sultana-Dyer solution and the solutions of Ref.~\cite{FJ} 
are perfectly comoving.

\section{Thermodynamics of a conformally  Schwarzschild 
cosmological black hole}

Let us consider now the zeroth law of black hole thermodynamics 
({\em i.e.}, the surface gravity $\kappa$ is constant on the 
horizon) 
for these cosmological black holes. In \cite{SD}, Sultana and 
Dyer assumed the temperature of their black hole solution to be
\begin{equation}\label{SDprescription}
T_{SD}= \frac{1}{2\pi} \left[ \kappa_{DH}-{\cal L}_{\xi} \left( 
\ln \Omega^2 \right) \right]=  \frac{1}{8\pi m_0} \;,
\end{equation}
{\em i.e.}, constant over the conformal Killing horizon and 
equal to the temperature of the static 
Schwarzschild black hole conformal to the Sultana-Dyer 
solution. Here ${\cal L}$ denotes the Lie derivative and 
$\xi^c$ is 
the conformal Killing vector which 
becomes null on the conformal Killing  horizon and satisfies
\begin{equation}
{\cal L}_{\xi}g_{cd}=\left( {\cal L}_{\xi} \ln \Omega^2 \right) 
g_{cd} \;,
\end{equation} 
while the surface gravity of the dynamical horizon is  defined 
by
\begin{equation} 
\xi^c \nabla_c \xi^a= - \kappa_{DH} \xi^a \;.
\end{equation}
Jacobson and Kang \cite{JacobsonKang}, instead, defined a 
generalized surface gravity $\kappa_{JK}$ defined by 
the normalization of the conformal Killing vector as 
\begin{equation} \label{JKsurfacegravity}
\nabla_a \left( \xi^c\xi_c \right)=-2\kappa_{JK}\xi_a \;.
\end{equation}
This is conformally invariant if $\Omega \rightarrow 1 $ at 
infinity. The relation between these two notions of surface 
gravity is 
\cite{JacobsonKang, SHM}
\begin{equation}
\kappa_{JK}=\kappa_{DH}-{\cal L}_{\xi}\left( \ln \Omega^2 
\right) \;,
\end{equation}
from which it follows that the corresponding  
temperatures for the cosmological black hole coincide,
\begin{equation} \label{wrongT}
T_{SD}=T_{JK}\equiv T_{JKSD}=\frac{1}{8\pi m_0} \;.
\end{equation}
However, this prescription for the black hole temperature is at 
odds with  generalizations of the zeroth law to conformal Killing 
horizons existing in metrics that are conformal to 
asymptotically flat  black holes 
\cite{DyerHonig,SDJMP}, and also with 
a simple argument proposed below. Saida, Harada, and Maeda 
argue that the black hole temperature should be the 
one given by the spectrum of the emitted Hawking radiation, 
which is instead \cite{SHM}
\begin{equation}
T_{SHM}=\frac{T_{JKSD}}{\Omega}=\frac{1}{8\pi m_0 a} \;.
\end{equation}
In the light of the previous discussion, this is simply $ 
T_{SHM}=1/\left(  
8\pi m_H \right)$, which appears natural when $m_H$ is regarded 
as the physical mass of the Sultana-Dyer black hole and 
reduces to the usual $T^{(Schw)}=\left( 8\pi m \right)^{-1}$ 
for a Schwarzschild black hole. This 
conclusion is supported by the following scaling argument. As 
discussed 
in great detail by Dicke \cite{Dicke}  following earlier 
ideas of Weyl \cite{Weyl}, a conformal transformation 
$g_{ab}\rightarrow \Omega^2 g_{ab}$ can be interpreted as a 
mere rescaling of the lengths of vectors and of the units used 
in a measurement, with the amount of rescaling depending  
on the spacetime position (although Dicke was concerned with 
the then-new Brans-Dicke theory \cite{BD}, his argument is 
quite general and applies also to general relativity as well 
as other metric theories of gravity). All that is measured in 
an 
experiment is the {\em ratio}  between a quantity $q$ and its 
unit $q_u$. For example, the proper length of a ruler  
divided by the unit of length $l_u $ is 
the same in the Minkowski 
metric $\eta_{ab}$ and in a conformally related metric 
$g_{ab}=\Omega^2 \eta_{ab}$ if a new length unit 
$\tilde{l}_u = \Omega l_u$ is associated to it---see 
Ref.~\cite{DickePeebles} for an application.  Therefore,  
two metrics $g_{ab}$ and $\tilde{g}_{ab}$ are physically 
equivalent \footnote{In modern language and in generalized 
gravity theories, this is known as the equivalence between the 
{\em 
Jordan conformal frame} and the {\em Einstein conformal frame} 
(for a recent discussion see Ref.~\cite{FaraoniNadeau}).}  
provided that the units of the fundamental 
quantities length, time, and energy   scale according to 
$\tilde{l}_u=\Omega\, l_u$, $ \tilde{t}_u=\Omega \, t_u$, and 
$\tilde{m}_u=\Omega^{-1} \, m_u$ \cite{Dicke} (derived units 
are scaled accordingly to their dimensions). In this sense, 
there is no difference between using the Schwarzschild metric 
$g_{ab}^{(Schw)}$ and its conformal Sultana-Dyer cousin 
$\tilde{g}_{ab}= g_{ab}^{(SD)}$, provided that the units 
$\tilde{l}_u$, $\tilde{t}_u$, and $\tilde{m}_u$ are 
appropriately scaled, {\em i.e.}, expanding for lengths and 
times, and redshifting away for energies. Since the black hole 
temperature $T$ 
multiplied by the Boltzmann constant $k_B$ scales as an energy,  
the ratio between $k_B T$ and $m_u$ must be the 
same 
when using $g_{ab}^{(Schw)}$ or $g_{ab}^{(SD)}$, or 
\begin{equation}
\frac{k_B \tilde{T }}{\tilde{m}_u}=
\frac{k_B T^{(Schw)}}{ m_u} \;,
\end{equation}
which yields the effective  temperature of the cosmological 
black hole
\begin{equation}\label{trueT}
\tilde{T}=\frac{T^{(Schw)}}{\Omega}=\frac{1}{8\pi m_0 
a}=\frac{1}{8\pi m_H}
\end{equation} in agreement with Ref.~\cite{SHM}. This simple 
argument supports the result of Saida, Harada, and 
Maeda \cite{SHM} and is fully consistent with the revealing 
use of the Hawking-Hayward quasi-local energy $m_H$ rather than 
other mass notions. 
The argument does not apply to a Schwarzschild-de Sitter (or 
Einstein-Straus) black hole, which can not be obtained by 
conformally transforming the Schwarzschild metric (remember that 
$m_H=$const. for this case, contrary to $m_H(t)=a(t) \, m_0$ for 
the Sultana-Dyer black hole).

In scalar-tensor cosmology it is well-known that simple 
rescaling provides the transformation law of the matter 
energy-momentum tensor under conformal transformations 
$g_{ab}\rightarrow 
\tilde{g}_{ab}=\Omega^2 g_{ab}$ as 
\begin{equation}
\tilde{T}^{(m)}_{ab}=\Omega^{-2}\, T^{(m)}_{ab} \;,
\end{equation}
which agrees with a direct calculation of 
$\tilde{T}^{(m)}_{ab}$ 
\cite{Wald, mybook}. By applying the rescaling to the 
semiclassical stress-energy tensor of a scalar field in the 
background of  a Sultana-Dyer (or any other comoving) black 
hole in our general-relativistic situation, the 
renormalized $\langle \tilde{T}_{ab} \rangle $ should then be
\begin{equation}
\langle \tilde{T}_{ab} \rangle =\frac{ \langle T_{ab} 
\rangle  }{a^2} \;.
\end{equation}
The explicit renormalization of $T_{ab}$ by Saida, 
Harada, and Maeda \cite{SHM} 
instead yields
\begin{equation} \label{trueset}
 \langle \tilde{T}_{ab} \rangle =
\langle T^{(SD)}_{ab} \rangle=
 \frac{ \langle T_{ab} 
\rangle  }{a^2} -\frac{1}{2880\pi^2}\left( X_{ab}-Y_{ab} 
\right) \;,
\end{equation}
where \footnote{Beware of the fact that tilded and non-tilded 
quantities are inverted in the notations of Ref.~\cite{SHM}.}
\begin{eqnarray}
X_{ab}&=& 2\tilde{\nabla}_a\tilde{\nabla}_b \tilde{R}
-2\tilde{g}_{ab}\tilde{\Box}\tilde{R}+\frac{\tilde{R}}{2}\tilde{g}_{ab}-2\tilde{R}\tilde{R}_{ab} 
\;, \\
&&\nonumber \\
Y_{ab}&=& -\tilde{R}_a^c \tilde{R}_{bc} 
+\frac{2}{3}\tilde{R}\tilde{R}_{ab}+\frac{1}{2} 
\tilde{R}_{cd} \tilde{R}^{cd} \tilde{g}_{ab} 
- \frac{\tilde{R}}{2} \tilde{g}_{ab} \;. \nonumber \\
&& 
\end{eqnarray}
The extra terms in eq.~(\ref{trueset}) are interpreted as due 
to quantum particle creation by the expanding background 
\cite{SHM}, which could not be predicted by using Dicke's 
classical argument. However, when the black hole 
horizon is 
much smaller than the cosmological horizon, these terms can be 
safely neglected and the rescaling argument agrees with the 
proper calculation of $\langle T^{(SD)}_{ab} \rangle $ in 
\cite{SHM}. 

An independent argument supporting the temperature~(\ref{trueT}) 
of cosmological black holes versus the expression~(\ref{wrongT}) 
is the following. It is instructive to consider the first law of 
black hole thermodynamics which, for a static Schwarzschild 
black hole of mass $m$ takes the form $ TdS=dm $. The expression 
of the Bekenstein-Hawking entropy $S= A/4$, where $A=4\pi 
r^2$ is  the horizon area, together with the expression $ 
r=2m$ for the 
horizon radius, yields the Hawking temperature 
 $  T^{(Schw)}=1/\left( 8\pi m \right) $. For a conformally 
expanding black hole of the Sultana-Dyer type or of the type in 
Ref.~\cite{FJ}, the quasi-local energy $m_H(t)=a(t)m(t)$ and 
proper 
horizon radius $r_p(t)=a(t) r$ (as well as proper area ${\cal 
A}=4\pi r_p^2$ and proper volume $ V=4\pi r_p^3/3$) should 
be used.  For these expanding horizons, the first law of black 
hole thermodynamics includes a work term $Pd V$:
\begin{equation}
TdS=dm_H+Pd V \;.
\end{equation}
By identifying again the black hole entropy with $S={\cal A}/4$ 
and using proper quantities, one obtains
\begin{equation}
8\pi T m_H dm_H=dm_H +32\pi P m_H^2 dm_H\;.
\end{equation}
In the adiabatic approximation  in which the accretion rate is 
small, the black hole is in a  state of quasi-equilibrium and the  
work term can be neglected yielding
\begin{equation}
T \simeq \frac{1}{8\pi m_H(t)}=\frac{T^{(Schw)}}{a} \;,
\end{equation}
in agreement with our previous argument. 
The disagreement between  the result of Ref.~\cite{SHM},  with 
which our arguments agree, and the 
Jacobson-Kang-Sultana-Dyer temperature is discussed in 
Ref.~\cite{SHM}. Recurrent folklore supports the idea that the 
black 
hole 
temperature is conformally invariant: however, the conformal 
invariance found in Ref.~\cite{JacobsonKang} is valid upon 
the assumption that the conformal factor satisfies $\Omega 
\rightarrow 1$ and that the conformal Killing field has unit 
norm at null infinity.  These assumptions are 
not satisfied by the Sultana-Dyer black hole \cite{SD}, 
nor by the  comoving black holes of Ref.~\cite{FJ}. The 
radiation spectrum can computed by evaluating 
the Bogoliubov coefficients relating ingoing and outgoing modes 
of positive and negative frequencies. The latter are defined by 
familiar boundary conditions when the spacetime is 
Minkowskian at infinity. These boundary conditions are not 
preserved by a  conformal transformation mapping an 
asymptotically flat spacetime into an 
asymptotically FLRW one (in this 
case the Bogoliubov coefficients are not 
expected to be conformally invariant). The temperature of these 
black holes in the adiabatic  approximation appears in 
Ref.~\cite{SHM} as  a result of a calculation of the  
renormalized energy-momentum tensor (a full semiclassical 
calculation including explicit Bogoliubov coefficients is not 
yet available).

From the physical point of view, it is clear that  the 
temperature of  an expanding black hole must be 
time-dependent while, if it were conformally 
invariant, it would be constant in time for a 
conformally-Schwarzschild black hole. In fact, $T$ is inversely 
proportional  to the physical mass; the latter {\em must} be 
related with the physical radius of the horizon ({\em e.g.}, 
by the expression of the Schwarzschild radius $r_s=2m$ for a 
Schwarzschild black hole). Therefore, since the horizon radius 
changes with time, also the physical mass changes with 
time, and so does the Hawking temperature. It would be 
unphysical for the temperature to remain time-independent while
the black hole expands  without bound.

\section{Static conformal transformation}

We now want to address an apparent contradiction 
\footnote{We thank a 
referee for bringing up this point.} 
between  the scaling argument proposed here and the claims of 
conformal invariance of the Hawking temperature appearing in the 
literature \cite{JacobsonKang, SD}. While this 
contradiction does not exist for the cosmological black hole 
considered so far because conformal invariance of the surface 
gravity and Hawking temperature has been demonstrated only for 
scale factors that approach unity at spatial infinity, it is 
certainly legitimate to consider a {\em stationary} conformal 
transformation in which the conformal factor does not depend on 
time and approaches unity at infinity. One can then consider the 
conformally transformed Schwarzschild black hole with, say, 
$\Omega=\Omega(r)$ in order to preserve spherical symmetry, and 
$\Omega\rightarrow 1 $ as $r\rightarrow +\infty$. The scaling 
argument still yields $\tilde{T}=\Omega^{-1} T$, in 
contradiction with the claim of conformal invariance 
$\tilde{T}=T$ \cite{JacobsonKang, SD}. This contradiction 
disappears 
when one realizes that two different notions of temperature are 
used, and that quasi-local energy and quasi-local mass behave 
differently under conformal transformations. In the following we 
adopt the Brown-York notions of quasi-local energy and mass 
\cite{BrownYork}. First, note that a stationary conformal 
transformation satisfies $ \chi ^c\nabla_c \Omega=0$, where 
$\chi^c$ 
is the timelike Killing vector of Schwarzschild spacetime, and 
therefore the Schwarzschild Killing horizon is mapped into 
another Killing horizon, not just a conformal Killing horizon. 
Second, the Brown-York expression for the quasi-local mass is 
conformally invariant \cite{BoseLohiya}: under general conformal 
transformations the latter is not a conserved charge, but it 
does enjoy this property for transformations with $ 
\chi^c\nabla_c 
\Omega=0$ \cite{BoseLohiya}. However, it is not the 
quasi-local 
mass that should be used here but rather the 
(Brown-York) quasi-local energy 
which differs 
from the quasi-local mass and has been used extensively in 
quasi-local black hole thermodynamics 
\cite{quasilocalbhthermodynamics}.  The boundaries 
that are necessary to define quasi-local quantities, in 
general, may not mapped into boundaries embeddable in the 
conformally related spacetime; however, this property holds for 
static, spherically symmetric, conformal transformations 
\cite{Capovillaetal}.

The quasi-local energy  $E$ is not 
conformally invariant but scales as  $\tilde{E}=\Omega^{-1}\, E$ 
\cite{BoseLohiya}. It is significant that, in the original 
Brown-York paper \cite{BrownYork}, the first law of 
thermodynamics applied to a Schwarzschild black hole with 
radius $R$ and mass $M$ becomes (eq.~(6.20) of 
Ref.~\cite{BrownYork})
\begin{equation}
\frac{dS}{8\pi M \sqrt{ 1-\frac{2M}{R} } } =dU+PdV \;,
\end{equation}
where $S=A/4=4\pi M^2$ is the entropy. The equilibrium 
temperature here is given by $ \left( 8\pi M \sqrt{ \left| 
g_{00}  \right| } \right)^{-1}$, not simply by $\left( 8\pi M 
\right)^{-1}$. This is 
reminiscent of Tolman's criterion for thermal equilibrium 
$T\sqrt{\left| g_{00} \right|}=$const. \cite{Tolman} which, 
applied to a 
Sultana-Dyer black hole, yields again $\tilde{T}=T/a$. For a  
stationary conformal transformation with $\Omega=\Omega(r)$, 
instead, if $M$ is conformally invariant, the new temperature 
will be 
\begin{equation}
\tilde{T}=\frac{1}{8\pi M \sqrt{ \left| \tilde{g}_{00} \right|} 
}=\frac{T}{\Omega} \;,
\end{equation}
which is consistent with the scaling relations
\begin{eqnarray}
d\tilde{U} &=&\frac{dU}{\Omega} \;,  \\
&&\nonumber \\
\tilde{P} &=&\Omega^{-4}P \;, \label{star}  \\
&&\nonumber \\
d\tilde{V} &=&\Omega^3 dV  \label{starstar} \;.
\end{eqnarray}
Eq.~(\ref{star}) is well known from the conformal 
transformation properties of perfect fluids in cosmology 
({\em e.g.}, Ref.~\cite{mybook}),  while eq.~(\ref{starstar}) 
applies to static 
conformal transformations for which $ d\Omega = 
\dot{\Omega} \, dt=0$ 
in any thermodynamic process. Therefore, the first law is valid 
also in the  conformally rescaled world and a necessary condition 
for this to happen is that $\tilde{T}=\Omega^{-1} T$. The 
contradiction between the scaling argument and the claimed 
conformal invariance originates from two different definitions of 
temperature, one based on the quasi-local energy 
\cite{BrownYork}, and the other  based on the conformally 
invariant surface gravity 
$\kappa_{JK}=\kappa_{DH}$ given by 
eq.~(\ref{JKsurfacegravity}) \cite{JacobsonKang, SD} and, in 
this respect, akin to the Brown-York quasi-local mass 

\section{Outlooks}

The previous considerations re-open the issue of which 
notion of temperature is to be used as the physical temperature 
of
a black hole that is conformally related to a static or 
stationary one or, more in general, of a dynamical horizon. We 
do not claim to have exhausted this subject here:  this issue  
is still open and awaits clarification.

To conclude, we have given independent arguments supporting the 
result  of \cite{SHM} for the temperature of a Sultana-Dyer 
black hole. A simple interpretation of this temperature is 
given, which appears particularly natural once 
the Hawking-Hayward quasi-local energy $m_H$ is adopted as the 
physical mass. The 
prescription~(\ref{trueT}) for the temperature of  
comoving cosmological black holes is extended to the solutions 
of Ref.~\cite{FJ}.

\begin{acknowledgments}
It is a pleasure to thank S. Sonego for a useful discussion in 
the Julian Alps.  
This work was supported by the  Natural Sciences and 
Engineering Research Council of Canada ({\em NSERC}). 
\end{acknowledgments}


\end{document}